\documentclass[a4paper,12pt]{elsarticle}
\DeclareGraphicsExtensions{.eps.pdf}
\usepackage{indentfirst}
\usepackage{amsthm}
\usepackage{rotating}
\begin{document}
\begin{frontmatter}
\author{Demetris P.K. Ghikas\corref{corr}}
\ead{ghikas@physics.upatras.gr}
\cortext[corr]{Corresponding author}
\author{Fotios D. Oikonomou}
\address{Department of Physics University of Patras,\\ Patras 26500, Greece}
\title{Towards an Information Geometric characterization/classification of Complex Systems. \\
II. Critical Parameter values from the (c,d)-manifold}

\begin{abstract}
In our previous paper (I) we derived information geometric objects from the two parameter generalized entropy of Hanel and Thurner (2011), using  the c,d parameters as labels of the corresponding manifolds. Here we follow a completely different approach by considering these parameters as coordinates of our information manifolds. This gives a manageable two-dimensional manifold amenable to easy manipulations but most importantly it offers a direct characterization of complex systems in terms of the pair of the c,d values. As a result we obtain certain characteristic values from the scalar curvature which we could conjecture that they represent complex systems with specific behavior. It is further observed that the boundary values of the c,d parameters which characterize the Hanel-Thurner classification are in some sense singular. This asks for a regularization scheme which we try to establish.   
\end{abstract}
\end{frontmatter}
\newpage
\section{Introduction}
\indent
\subsection*{Classification of complex systems and Generalized Entropies}
In our previous paper (I) \cite{ghi1} we have discussed both the problems of definition, characterization and classification of complex systems as well as the use of generalized entropies giving the most relevant references. Here our starting point is again the two-parameter distribution related to the (c,d)-entropy of Hanel and Thurner  \cite{hanel1} but we consider these parameters as coordinates of our information manifold. This offers a direct association of complex systems with particular values of these parameters where the scalar curvature is extremized. 

\subsection*{Information Manifolds from Probability Distributions}
Information Geometry emerged as a practical geometric framework in the theory of parameter estimation in mathematical statistics \cite{amari1}. For a given statistical model, that is a given class of probability measures there is associated an information manifold and a Riemannian metric. This metric enters in the estimation procedure through the Cramer-Rao Inequality giving the possible accuracy of an estimator of the parameters of the model. Further on, one may define non-Riemannian connections which offer a deeper analysis of the estimation procedure. Our work is based on the geometric quantities emerging in the Information Geometry which is based on the two-parameter entropy functional of Hanel and Thurner \cite{hanel1}. In this paper the Hanel-Thurner parameters are used as coordinates of the information manifold. Thus we obtain a two-dimensional manifold without any approximation, and we may use both analytical and numerical results to establish the structure of this. 
\\
\indent
In Paragraph 2 we introduce in a minimal way the necessary definitions and geometric quantities of Information Geometry.  In Paragraph 3  the generalized entropy of Hanel and Thurner \cite{hanel1} is introduced with few comments on its derivation and properties. In Paragraph 4 we present our results. First we state some properties which establish the appropriateness of the generalized distribution function. Then we compute the Riemannian metric for the (c,d)-distribution function and the scalar curvature. The first three graphs give aspects of the distribution function. The next three  present the metric elements as functions of the coordinates c,d giving the first impression of their non-trivial form. Then the numerical construction of the scalar curvature is given in the seventh graph. From this it is evident that some areas of the manifold are not typical. To reveal this structure  we present certain sections of the curvature surface as functions of the c parameter for some values of the d parameter. We also give the contour plots of the curvature and the locus of minima as a function of c and d. We could conjecture that these extreme values correspond to complex systems of particular properties. 
  
\section{Basic concepts of Information Geometry}
\subsection*{Geometry from probability distributions and the Cramer-Rao Inequality}
\indent 
Here we present only the necessary concepts in order to establish the notation. We refer to the bibliography for the details \cite{amari1}. Let
\begin{equation}
S=\{p_{\xi}=p(x;\xi) | \xi = [\xi^{1},...,\xi^{n}]\in \Xi\}
\end{equation}
be a parametric family of probability distributions on $\mathcal{X}$. This is an n-dimensional parametric statistical model. Given the N observations $x_{1},...,x_{N}$ the Classical Estimation Problem concerns the statistical methods that may be used to detect the true distribution, that is to estimate the parameters $\xi$. To this purpose, an appropriate estimator is used for each parameter. These estimators are maps from the parameter space to the space of the random variables of the model. The quality of the estimation is measured by the variance -covariance matrix $V_{\hat{\xi}}=[v_{\xi}^{ij}]$ where
\begin{equation}
v_{\xi}^{ij}= E_{\xi}[(\hat{\xi^{i}}(X)-\xi^{i})(\hat{\xi^{j}}(X)-\xi^{j})]
\end{equation} 
Suppose that the estimators are unbiased, namely
\begin{equation}
E_{\xi}[\hat{\xi}(X)]=\xi , \quad \forall \xi \in \Xi
\end{equation}
Then a  lower bound for the estimation error is given by the Cramer-Rao inequality
\begin{equation}
V_{\xi}(\hat{\xi})\geq G(\xi)^{-1}
\end{equation}
where $G(\xi)=[g_{ij}(\xi)]$ 
\begin{equation}
g_{ij}(\xi)=E_{\xi}[\partial_{i}l(x;\xi)\partial_{j}l(x;\xi)]
\end{equation}
the Classical Fisher Matrix with 
\begin{equation}
l_{\xi}=l(x;\xi)=\ln p(x;\xi)
\end{equation}
the score function.
As it has been shown the Fisher Matrix provides a metric on the manifold of classical probability distributions. This metric, according to the theorem of Cencov \cite{cencov}, is the unique metric which is monotone under the transformations of the statistical model. This means that if the map $F : \mathcal{X} \to \mathcal{Y}$ induces a model $S_{F}=\{q(y;\xi)\}$ on $\mathcal{Y}$ then 
\begin{equation}
G_{F}(\xi)\leq G(\xi)
\end{equation} 
That is, the distance of the transformed distributions is smaller than the original distributions. Thus monotonicity of the metric is intuitively related to the fact that in general we loose distinguishability of the distributions from any transformation of the information.
\newline
\indent
The metric defined in this way is the ordinary Fisher metric. Using the Levi-Civita connection the corresponding Riemannian structure is  constructed. In this geometry the scalar curvature is a quantification of the information manifolds. But there is a further development connected with the existence of connections different from Levi-Civita. These are certain pairs of connections satisfying a duality property with respect to the Fisher metric and playing a fundamental role in the estimation theory. An important case is the dually flat connections. 
\newline
\indent

\section{Generalized Entropies and Complex Systems}
\indent
\subsection{Generalized Entropies}
\indent
Assuming the four Shannon-Khinchin axioms it is proved that there exists a unique entropy functional, the Boltzmann-Gibbs Entropy

\begin{equation}
S[p] = -\sum_{j\in J}p(j)\ln p(j) \quad , \quad  \sum_{j\in J}p(j) = 1
\end{equation}
These axioms are plausible assumptions abstracted from the typical behavior of thermodynamic systems and the role of thermodynamic entropy. But after the statistical foundation of thermodynamics and the association of entropy with information theory, it became necessary to look for other functionals which were thought to cover more general systems than the simple ones like perfect gases, and more generally systems with long range interactions. And though it is expected that in the thermodynamic limit to have functionals with a universal form, it is evident that for small systems one needs functionals dependent on parameters. These parameters, not having always a transparent connection with the empirical properties of the systems, nevertheless, offered a minimal parametric generalization of the Boltzmann-Gibbs functional as an information theoretic tool. One of the earliest generalizations is the Renyi's Entropy

\begin{equation}
S^{q}[p] = \frac{1}{1-q}\ln ( \sum_{j}p(j)^{q} )
\end{equation}   
Later on Tsallis \cite{tsall1,tsall2,tsall3}, in relation to the theory and practice of fractals introduced his entropy

\begin{equation}
S_{q}^{Tsallis}[p] = \frac{1}{1-q}( \sum_{j}p(j)^{q}-1 )
\end{equation}
a form that had been introduced earlier for mathematical reasons. Thereafter a host of other forms of entropic functionals were introduced associated with particular properties of complex statistical systems. All these entropies, assuming a form of Maximal Entropy Principle give rise to probability distributions which depend on the parameter of entropy. In general these are generalized exponentials which are the inverse functions of generalized logarithms. These generalized exponentials, assumed to be particular exponentials of probability distributions may be used to construct information geometric objects. In this work we use the two-parameter entropic functional of Hanel and Thurner to construct our geometric tools. 
 
\subsection{A two-parameter Generalized Entropy and Complex Systems}
\indent

Given the fact that the four Shannon-Khinchin Axioms impose a unique form for the entropy, which is the Boltzmann-Gibbs functional, Hanel and Thurner, seeking a generalization to the case of a functional not satisfying additivity had to abandon the relevant axiom. Their analysis produced a two-parameter entropic functional of the form

\begin{equation}
S_{c,d}[p] =  \frac{e\sum_{i}^{W}\Gamma(d+1,1-c\ln p_{i})}{1-c+cd} - \frac{c}{1-c+cd}
\end{equation}
where W is the number of potential outcomes and $\Gamma(a,b) = \int_{b}^{\infty}dt t^{a-1}exp(-t)$ the incomplete Gamma-function.
 The Bolzmann-Gibbs entropy is recovered for (c,d) = (1,1), while for the Tsallis entropy we have (c,d) = (c,0).
\newline
\indent
The associated distribution function is the generalized exponential
\begin{equation}
\mathcal{E}_{c,d,r}(x) = exp\left(-\frac{d}{1-c}[W_{k}(B(1-x/r)^\frac{1}{d}) - W_{k}(B)]\right) 
\end{equation}
where $r = (1-c+cd)^{-1}$ and $B = \frac{(1-c)r}{1-(1-c)r}\exp(\frac{(1-c)r}{1-(1-c)r})$. The function $W_{k}$ is the k-th branch of the Lambert W function which is a solution of the equation $x = W(x)exp(W(x))$. This generalized exponential is the inverse function of the generalized logarithm 
\cite{hanel1,hanel2,hanel3,hanel4,hanel5}.

\begin{equation}
\Lambda_{c,d,r}(x) = r-rx^{c-1}[1-\frac{1-(1-c)r}{rd}\ln x ]^{d}
\end{equation}

\section{Results}

\subsection*{The (c,d)-manifold}
\indent
In our previous paper (I) \cite{ghi1} we considered c and d as parameters characterizing the manifolds derived from the (c,d)-exponential family. This gave us a certain classification scheme which characterized various classes of complex systems. Here our starting point is to consider a two-dimensional manifold with coordinates the parameters c and d. The normalized distribution is
\begin{equation}
p(c,d,x) = -\frac{c(1+c(d-1))\exp\left(d\left(W(L)+\frac{W\left(L(1+(c-1-cd)x)^{1/d}\right)}{c-1}\right)\right)}{(c-1)\exp\left(\frac{cdW(L)}{c-1} \right)-dE_{1-d}\left(-\frac{cd}{c-1}W(L)  \right)}
\end{equation}
for $0<c<1$ , $0<d<1$ and $-\infty <x\leq 0$ where 
\begin{equation}
L = L(c,d) = -\frac{(c-1)\exp\left(\frac{1-c}{cd}\right)}{cd} = B\left(c,\frac{1}{1-c+cd}\right)
\end{equation}
and
\begin{equation}
E_{n}(z) = \int_{1}^{+\infty}\frac{\exp(-zt)}{t^{n}}dt
\end{equation}
From this
we computed the Fisher metric and  the scalar curvature. Unfortunately the computations may be performed only numerically. We present the resulting metric elements and  the scalar curvature.  From these plots its is evident first that the metric is a non trivial function of c and d. And most impressively, that the scalar curvature has at certain regions of the (c,d)-manifold strong variations. To investigate these variations we made cross sections which give for various values of d , c-dependent curves. These curves have very steep minima and some maxima. We complement this picture with contour plots of the curvature and  a graph  with the locus of the minima in the c,d plane. We dare to conjecture that these extrema are associated to characteristic values for the corresponding complex systems.

\begin{figure}[htp]
	\centering
\scalebox{0.8}{\includegraphics[angle=0]{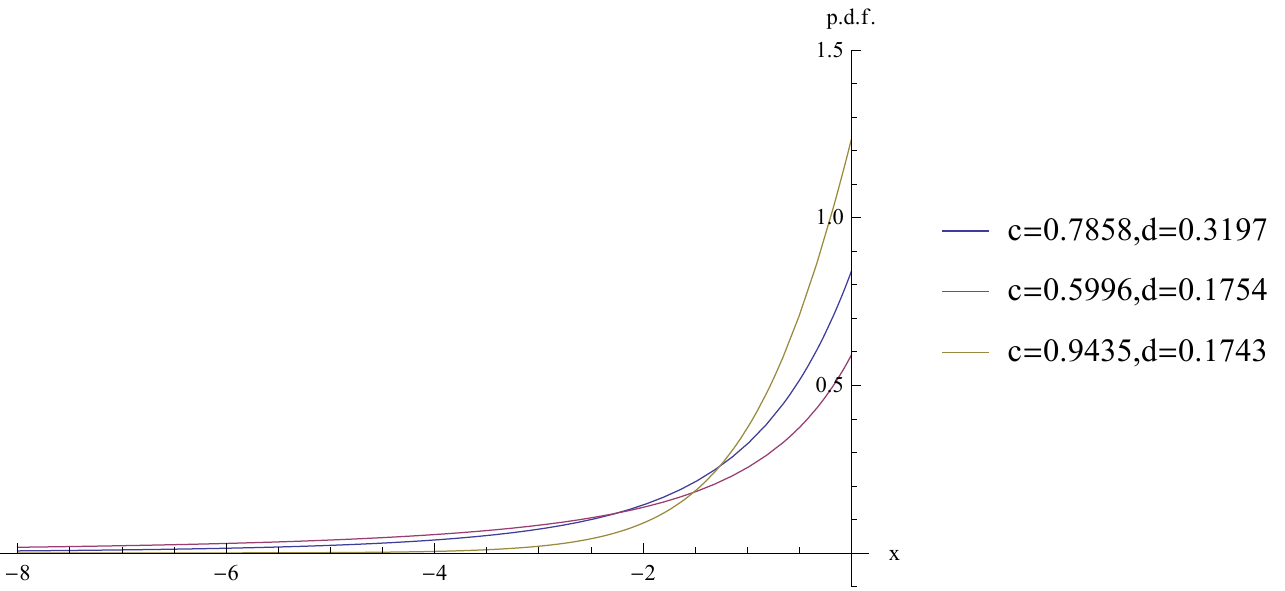}}
\caption{Normalized Probability Density}
	\label{fig: pdf1}
\end{figure}

\begin{figure}[htp]
	\centering
\scalebox{0.8}{\includegraphics[angle=0]{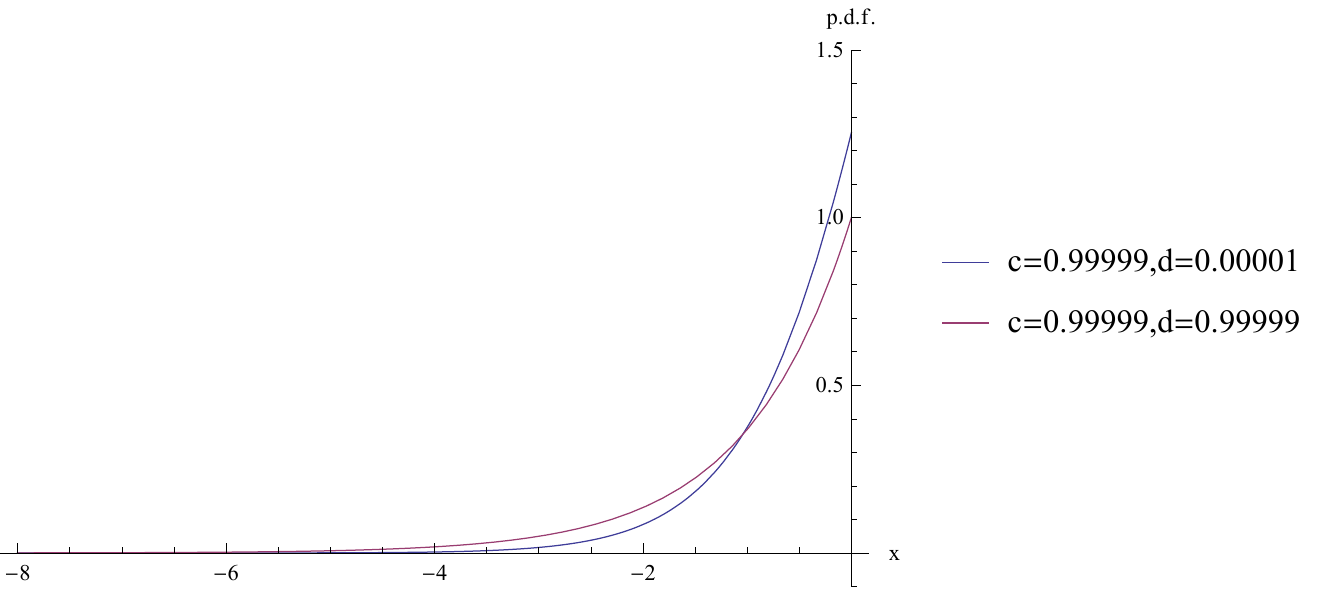}}
\caption{Normalized Probability Density}
	\label{fig: pdf2}
\end{figure}

\begin{figure}[htp]
	\centering
\scalebox{0.8}{\includegraphics[angle=0]{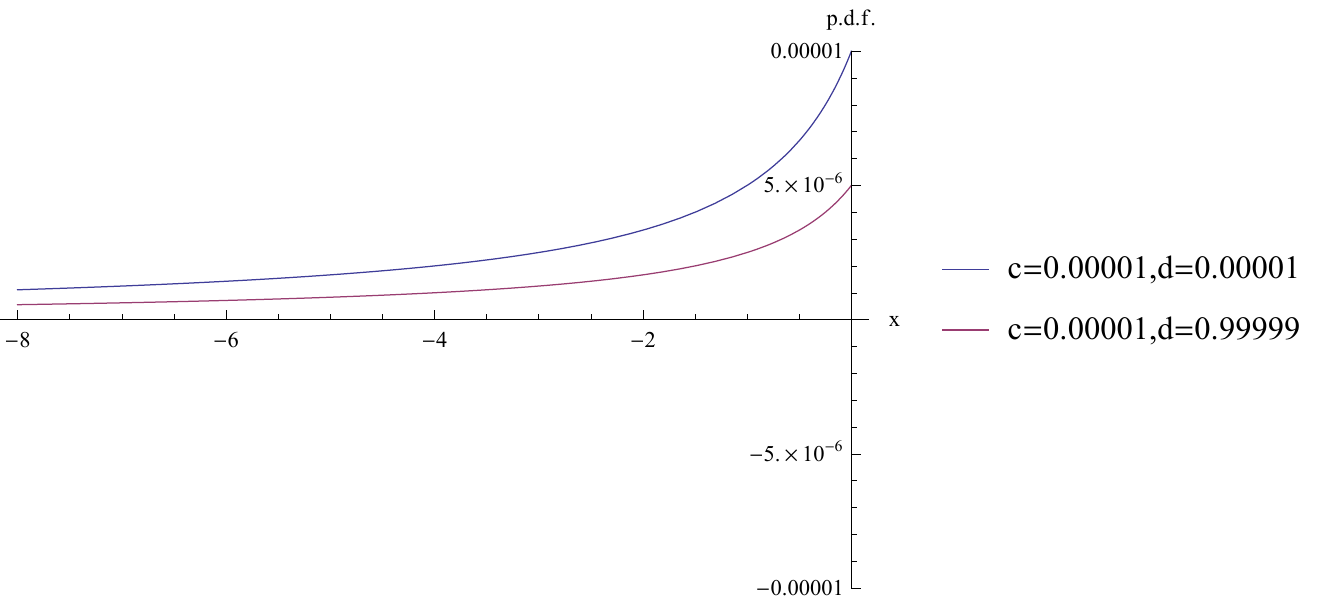}}
\caption{Normalized Probability Density}
	\label{fig: pdf3}
\end{figure}

\begin{figure}[htp]
	\centering
\scalebox{0.8}{\includegraphics[angle=0]{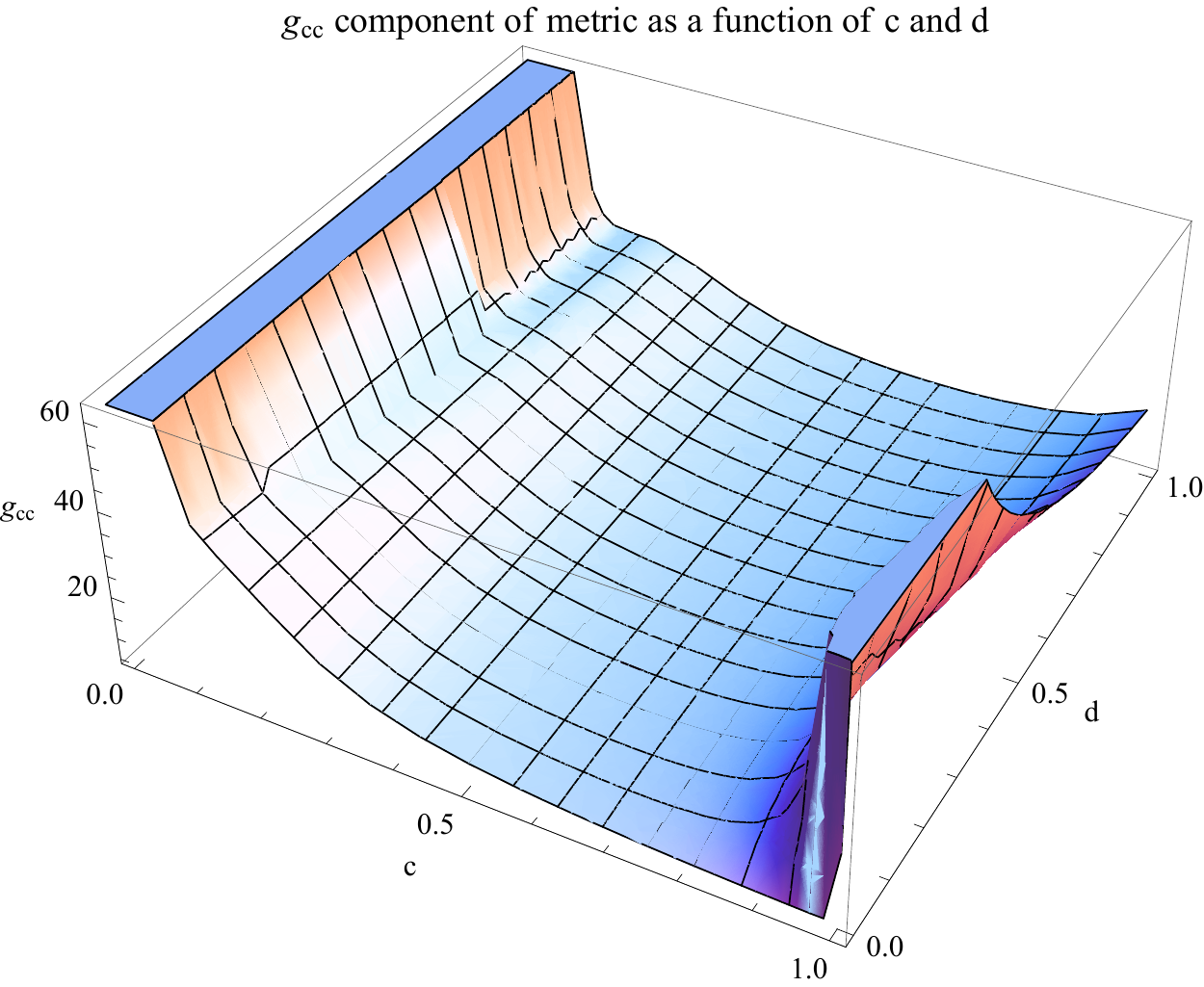}}
\caption{metric element $g_{11}$}
	\label{fig: gcc}
\end{figure}

\begin{figure}[htp]
	\centering
\scalebox{0.8}{\includegraphics[angle=0]{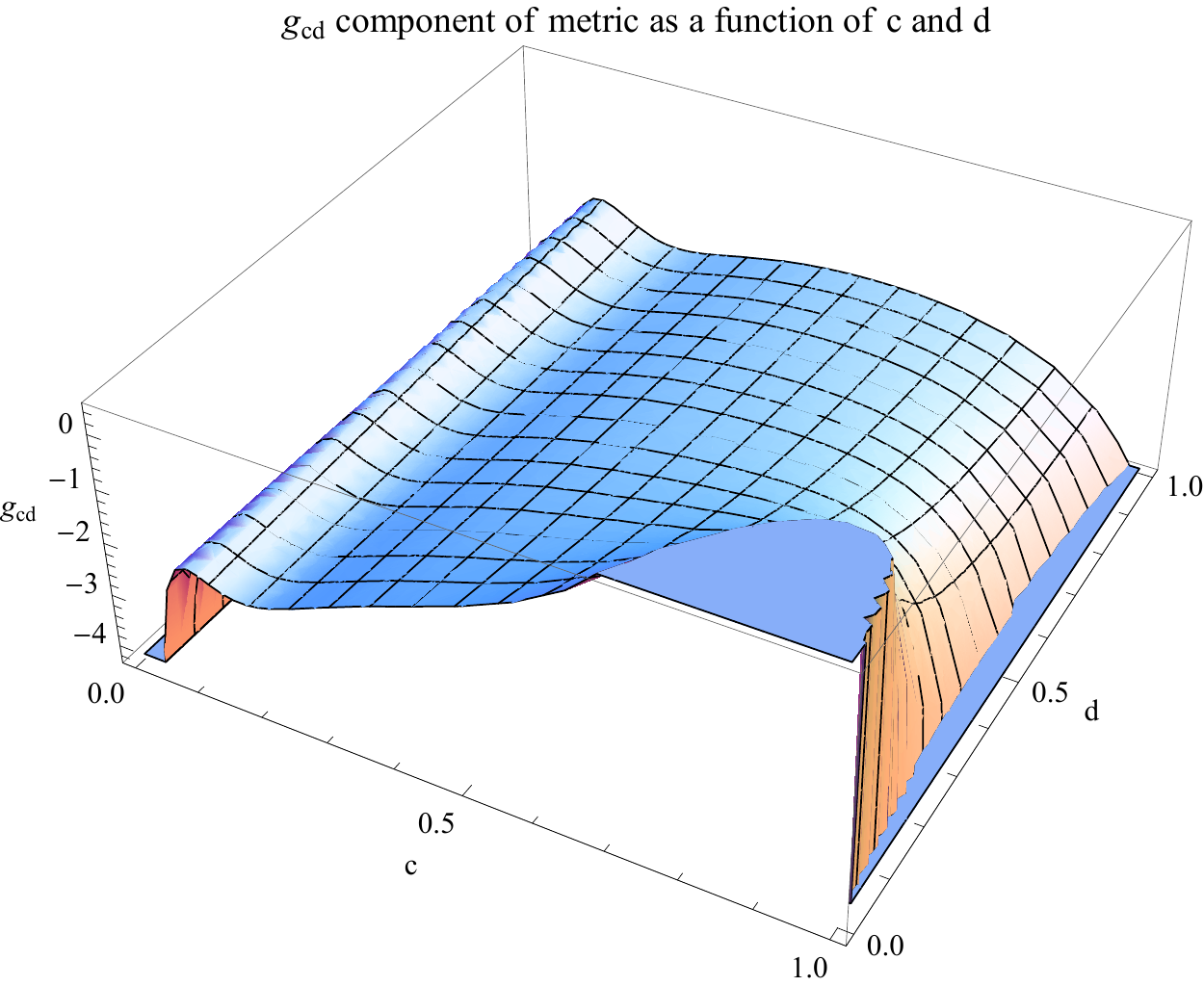}}
\caption{metric element $g_{12}$}
	\label{fig:gcd}
\end{figure}

\begin{figure}[htp]
	\centering
\scalebox{0.8}{\includegraphics[angle=0]{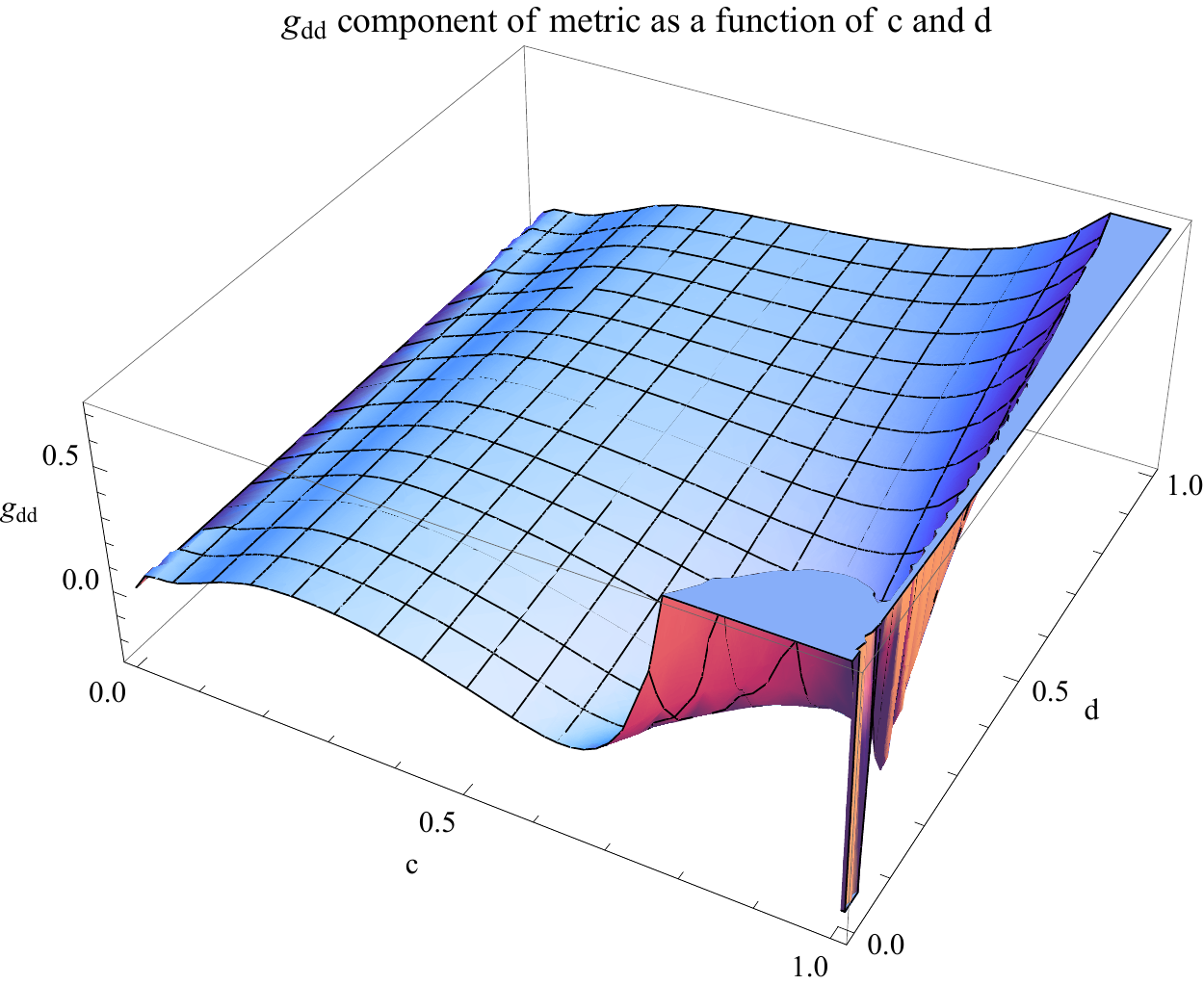}}
\caption{metric element $g_{22}$}
	\label{fig:gdd}
\end{figure}

\begin{figure}[htp]
	\centering
\scalebox{0.8}{\includegraphics[angle=0]{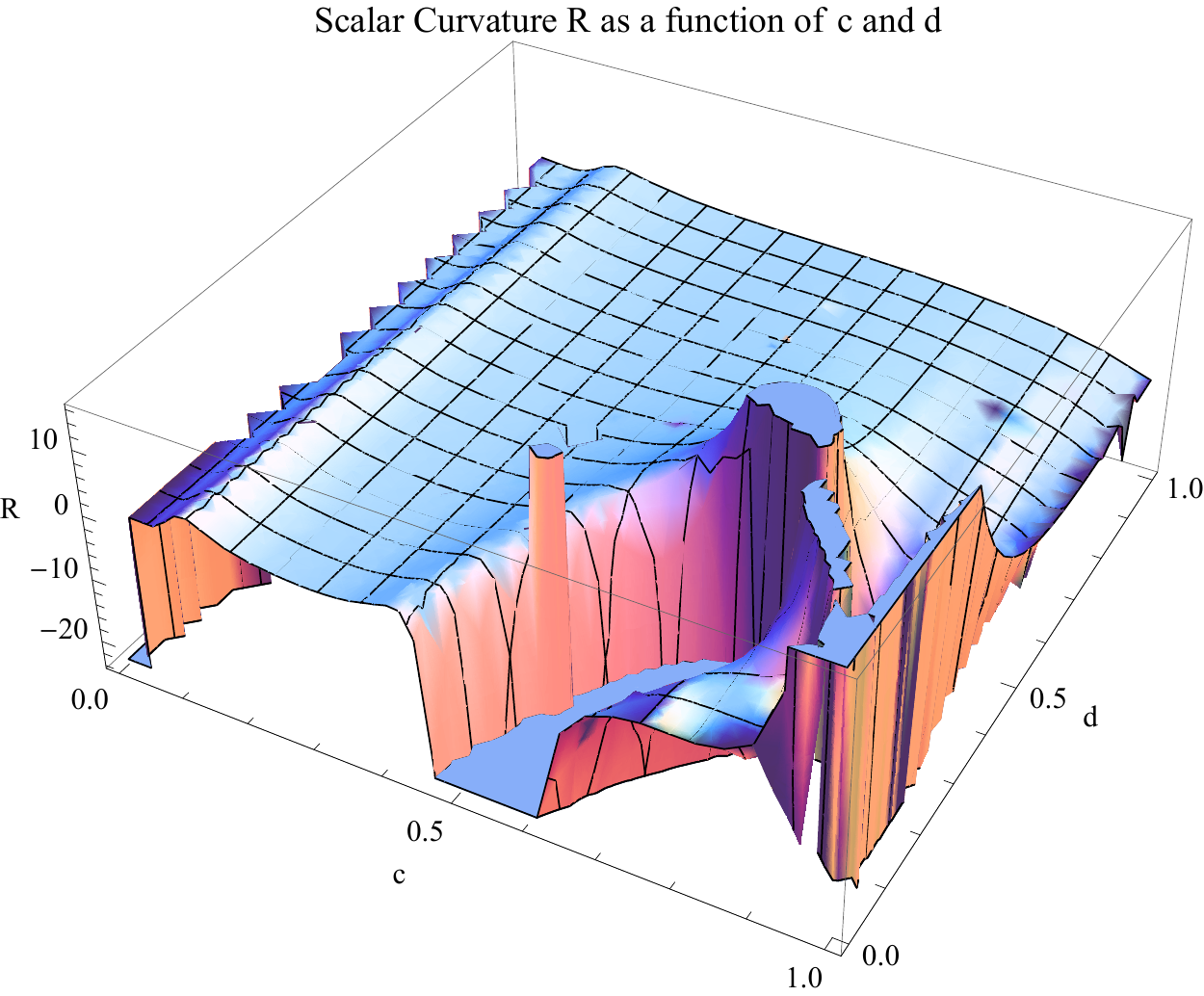}}
\caption{Scalar Curvature}
	\label{fig:scalar}
\end{figure}

\begin{figure}[htp]
	\centering
\scalebox{0.8}{\includegraphics[angle=0]{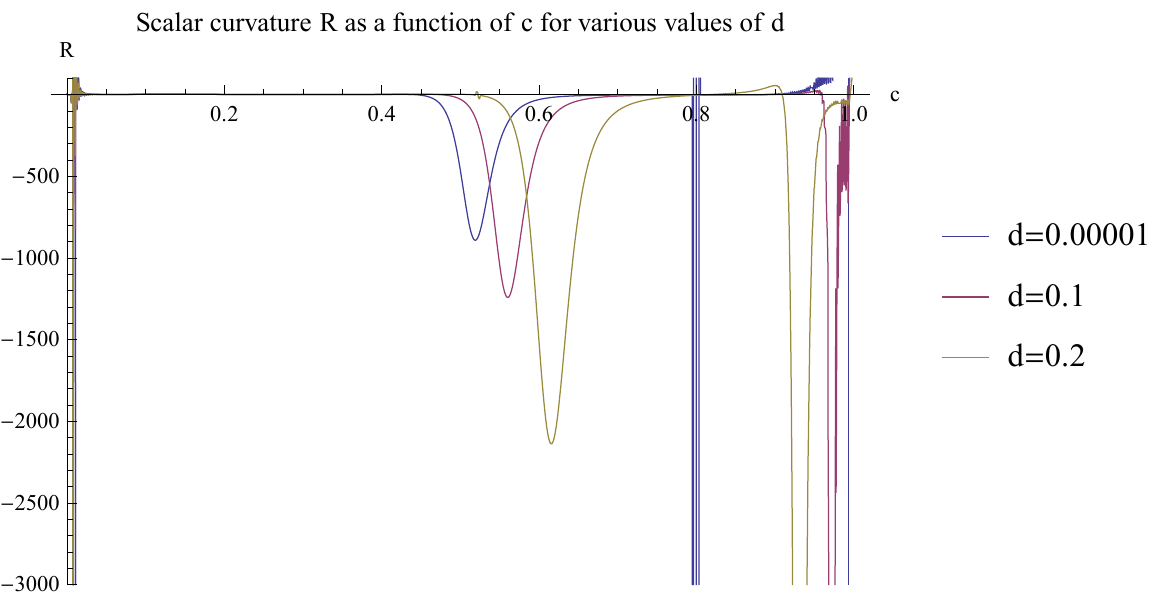}}
\caption{Section of Scalar Curvature 1}
	\label{fig: section 1}
\end{figure}

\begin{figure}[htp]
	\centering
\scalebox{0.8}{\includegraphics[angle=0]{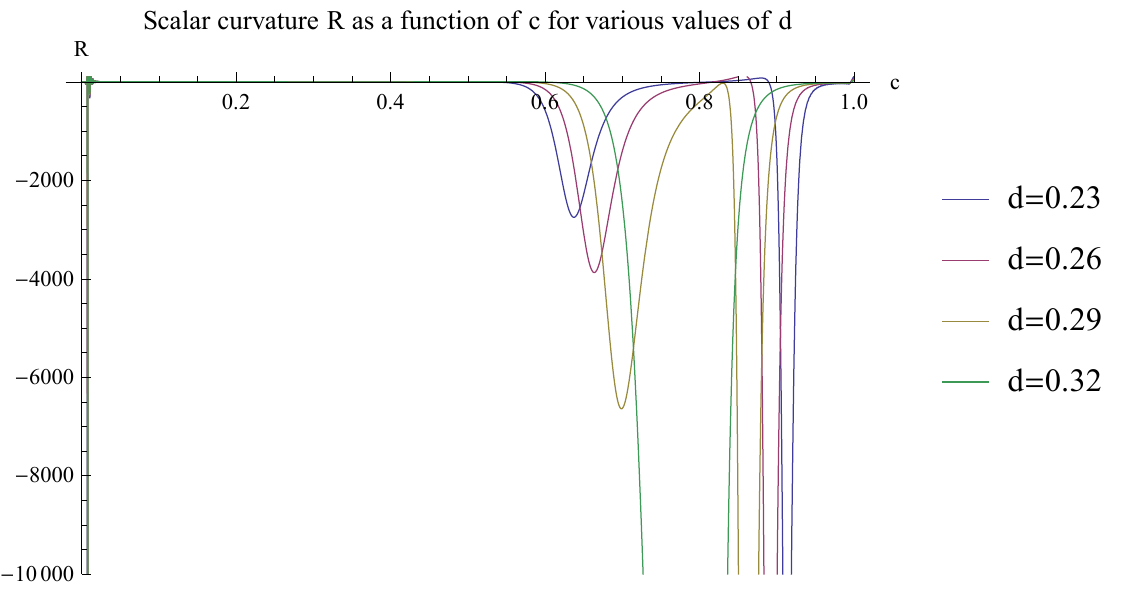}}
\caption{Section of Scalar Curvature 2}
	\label{fig:section 2}
\end{figure}

\begin{figure}[htp]
	\centering
\scalebox{0.8}{\includegraphics[angle=0]{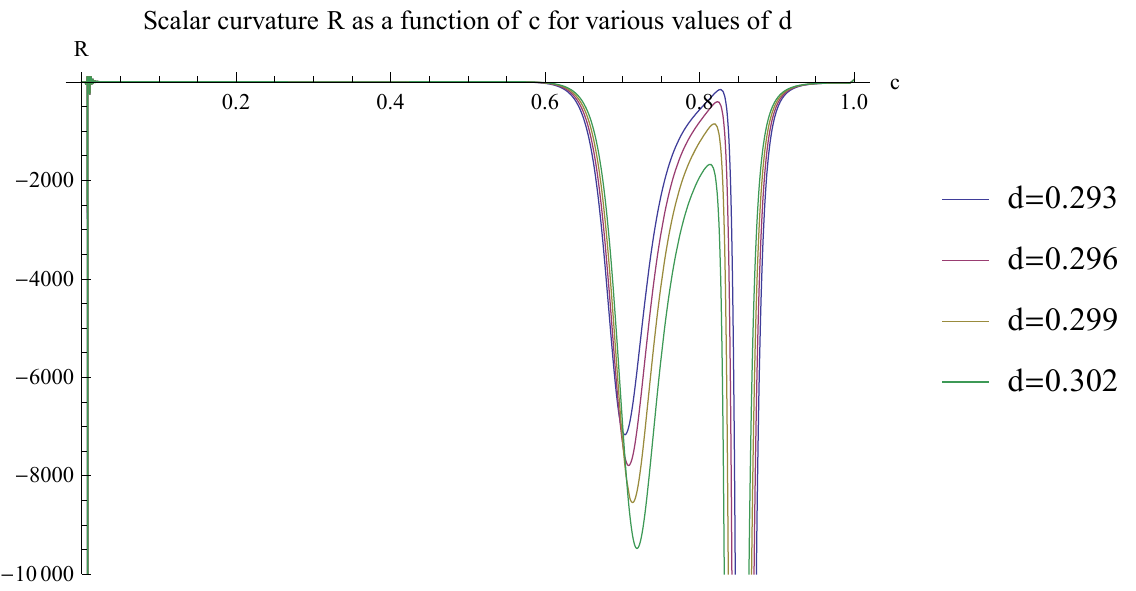}}
\caption{Section of Scalar Curvature 3}
	\label{fig:section 3}
\end{figure}

\begin{figure}[htp]
	\centering
\scalebox{0.8}{\includegraphics[angle=0]{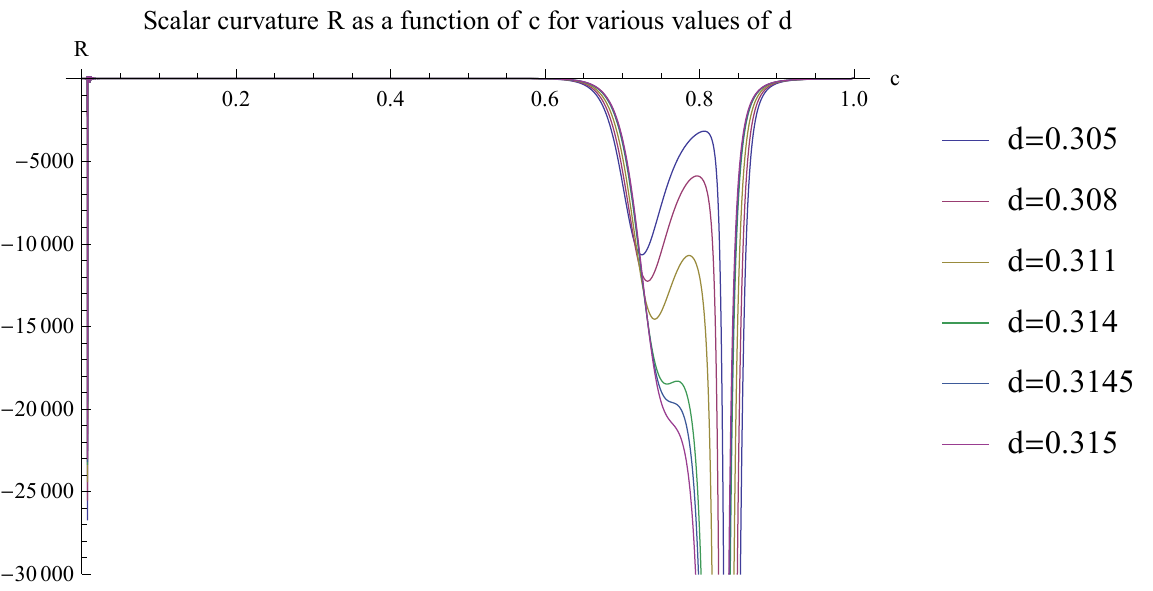}}
\caption{Section of Scalar Curvature 4}
	\label{fig:section 4}
\end{figure}

\begin{figure}[htp]
	\centering
\scalebox{0.8}{\includegraphics[angle=0]{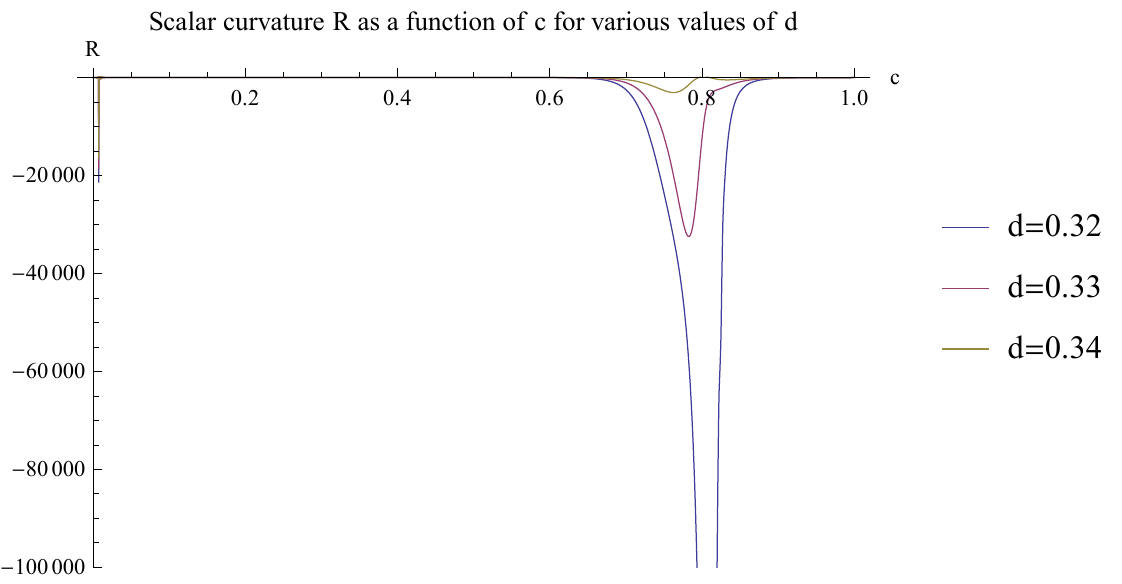}}
\caption{Section of Scalar Curvature 5}
	\label{fig:section 5}
\end{figure}

\begin{figure}[htp]
	\centering
\scalebox{0.8}{\includegraphics[angle=0]{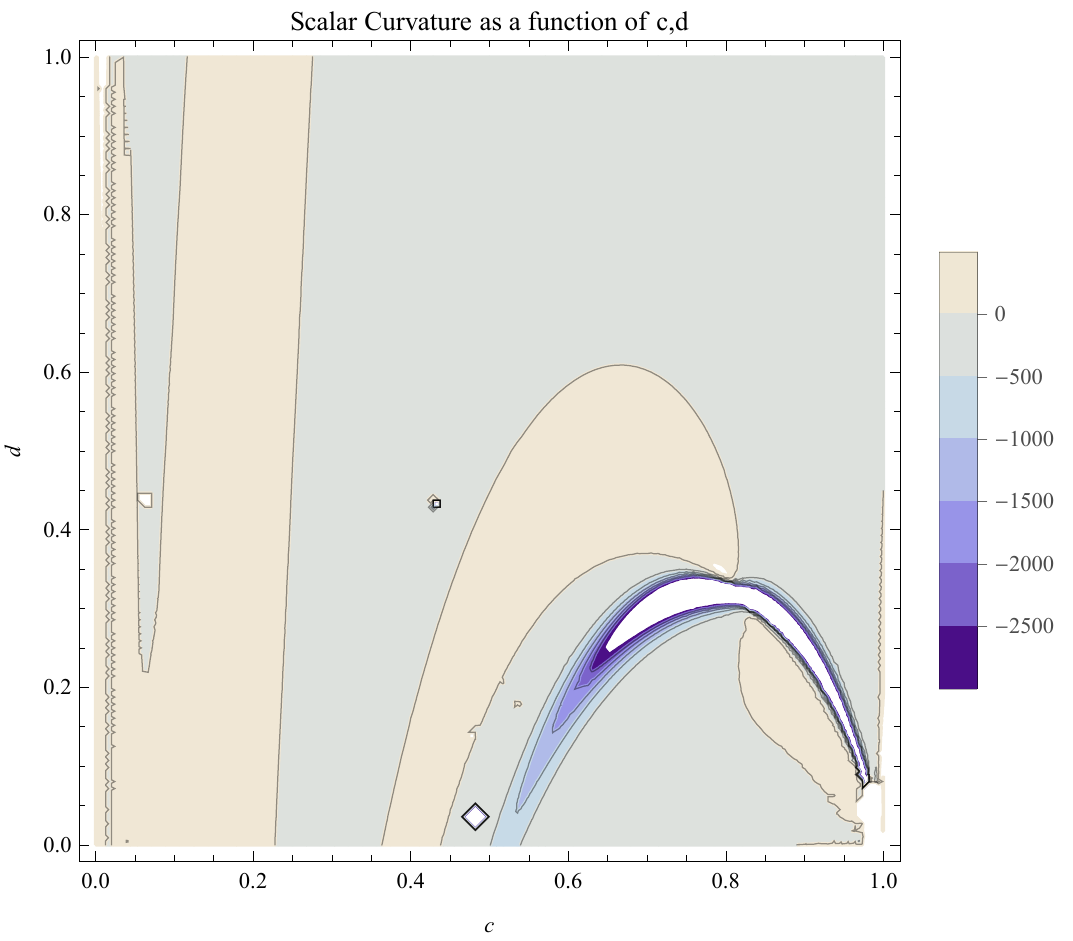}}
\caption{Exrema of Scalar Curvature}
	\label{fig:extrema}
\end{figure}

\begin{figure}[htp]
	\centering
\scalebox{0.8}{\includegraphics[angle=0]{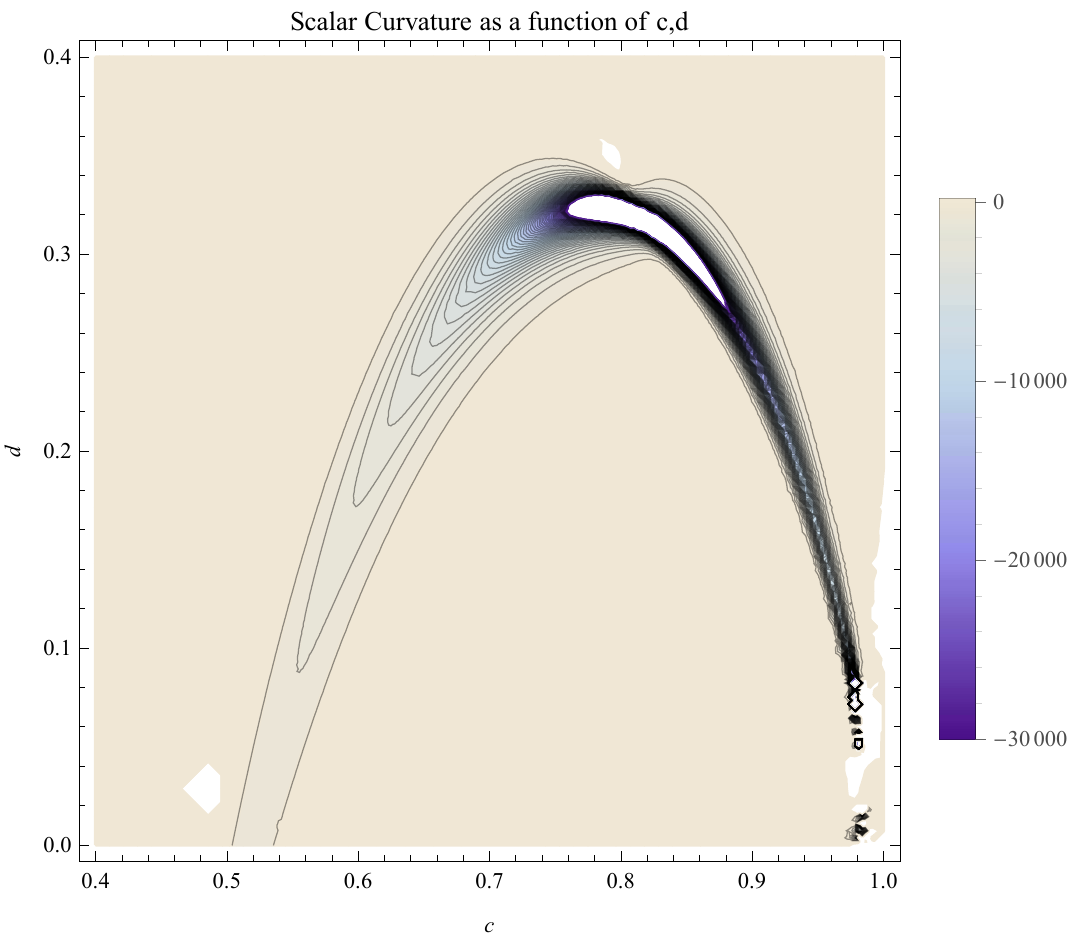}}
\caption{Zoom of Extrema of Scalar Curvature}
	\label{fig:zoom extrema}
\end{figure}

\begin{figure}[htp]
	\centering
\scalebox{0.8}{\includegraphics[angle=0]{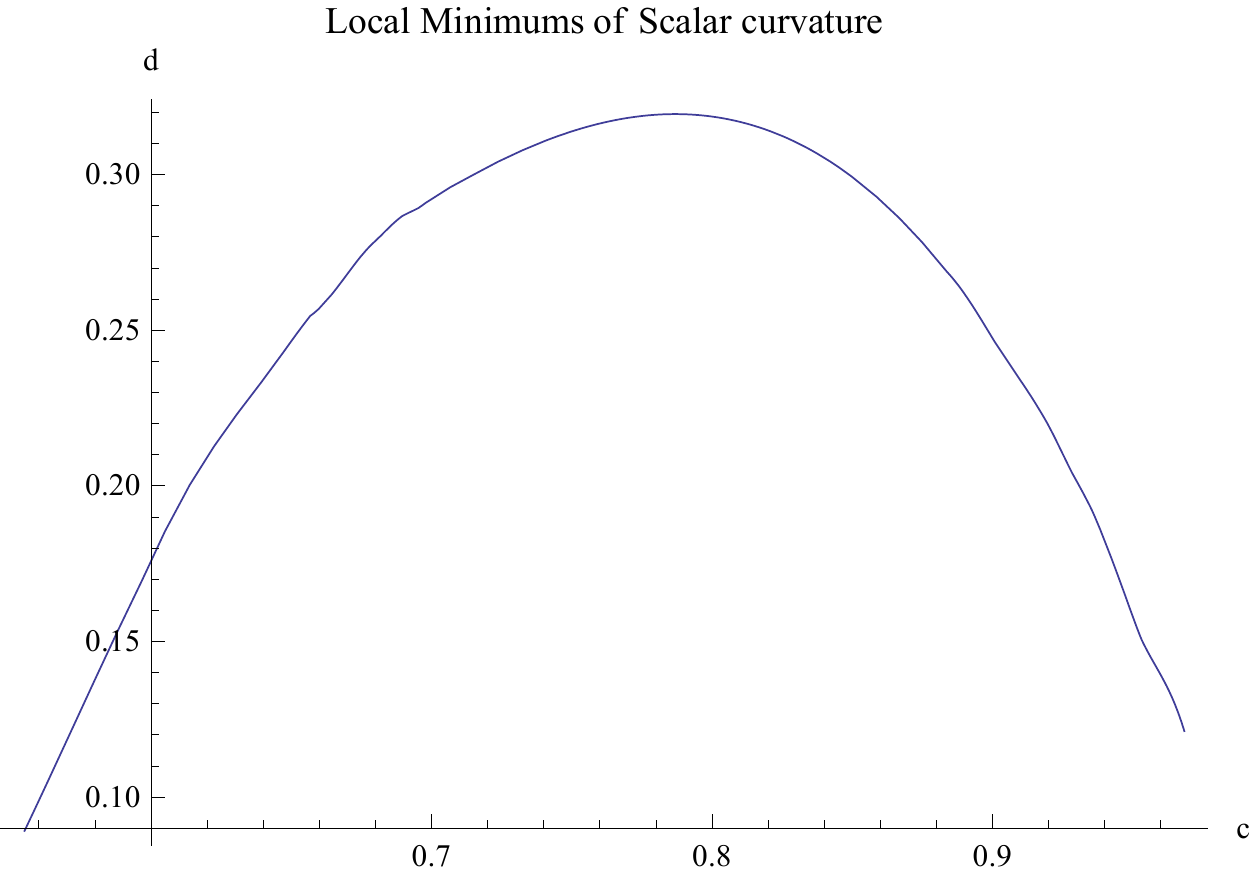}}
\caption{Local Minima  of Scalar Curvature}
	\label{fig:minima extrema}
\end{figure}

\section{Discussion}
\indent
In our previous paper (I) \cite{ghi1} we posed the question whether the generalized (c,d)-entropy of Hanel and Thurner \cite{hanel1,hanel2,hanel3,hanel4,hanel5}, and the corresponding generalized exponential may give information manifolds that show clear dependence on these parameters. Indeed , various geometric properties derived from the information manifolds clearly classified them in five classes corresponding to the classes associated to the extreme values of the c,d parameters. Here we change completely our point of view. We consider the c,d parameters not as parameters of the information manifolds but as coordinates of a single information manifold. In this case the boundary values correspond to the Hanel-Thurner classes, but we have a landscape of pairs of c,d parameters where complex systems should correspond. The manifold is two-dimensional and theoretically easier to handle. Unfortunately, due to the complicated form of the distribution calculations may be done only numerically. Thus we numerically constructed the metric and the scalar curvature of this 2d information manifold. The metric shows that we have a nontrivial geometry. But the scalar curvature revealed the existence of regions in the c,d plane where some dramatic variations exist. To exhibit more clearly these variations we present sections of the curvature plot as curves depending on c for various values of d. In these sections we see that we have some extreme values of the curvature, minima and some maxima. Drawing contour plots we reveal the special regions in the c,d plane. We estimated the locus of the minima and we got  a smooth curve which gives pairs of values of the c,d parameters where it seems that they correspond to complex systems with special properties. It is tempting to conjecture that these systems do exist. 
\\
\indent
Finishing we must point to a serious difficulty concerning the boundary c,d values. We may not approach them numerically. We need some kind of regularization and smooth matching to the boundary values which correspond to various classes of systems. We are currently investigating this problem. We are also looking to the Cramer-Rao inequalities near the boundary values and along the curves of extrema. These inequalities are related to estimation theory but we have to understand how the c,d values are related to measurable quantities.  
\section*{Acknowledgments}
One of the authors (F.O.) wishes to thank the Greek State Scholarships Foundation for a research scholarship.
\newline

\end{document}